# Gate-tunable memristors from monolayer MoS$_2$

Vinod K. Sangwan[1,*], Hong-Sub Lee[1,*], and Mark C. Hersam[1,2,3,+]

[1]Department of Materials Science and Engineering, [2]Department of Chemistry, [3]Department of Electrical Engineering and Computer Science, Northwestern University, Evanston, Illinois 60208, USA.
[*]These authors contributed equally; [+]Corresponding author: m-hersam@northwestern.edu

*Abstract*—We report here gate-tunable memristors based on monolayer MoS$_2$ grown by chemical vapor deposition (CVD). These memristors are fabricated in a field-effect geometry with the channel consisting of polycrystalline MoS$_2$ films with grain sizes of 3-5 μm. The device characteristics show switching ratios up to ~500, with the resistance in individual states being continuously gate-tunable by over three orders of magnitude. The resistive switching results from dynamically varying threshold voltage and Schottky barrier heights, whose underlying physical mechanism appears to be vacancy migration and/or charge trapping. Top-gated devices achieve reversible tuning of threshold voltage, with potential utility in non-volatile memory or neuromorphic architectures.

## I. INTRODUCTION

Recently, two-terminal memristors have emerged as a potential replacement for flash memory and as a foundational circuit element for neuromorphic computing and non-Boolean logic architectures [1-6]. Memristor-based resistive random access memory (ReRAM) shows lower programming voltage, smaller footprint, faster read/write time, and higher endurance than NOR and NAND flash memory [7]. However, cross-point architectures based on two-terminal memristors either involve an additional transistor for each cell or a complicated biasing scheme that limits scalability [7].

Two-terminal memristors have also demonstrated basic neuromorphic functions such as potentiation and spike-timing-dependent plasticity [3,4]. Nevertheless, many synaptic activities of neurons also exhibit dependencies on a third modulatory terminal that cannot be fully captured by two-terminal memristors [8]. Several approaches have attempted to overcome these limitations through complementary switching in head-to-tail memristor architectures, cationic synaptic emulators, battery-like synapses, and three-terminal memristors [9-14]. While these methods address some issues, a universal neuromorphic architecture based on conventional metal oxide filamentary memristors remains elusive, suggesting a need for alternative memristor materials and geometries. Recently, two-dimensional materials such as MoS$_2$ have shown promise for memristor technology [15,16]. For example, gate-tunable memristive phenomena have been correlated with field-driven grain boundary motion in monolayer MoS$_2$ [16]. Here, we build off those early results and present a scalable memristor design that achieves gate tunability of resistance states in polycrystalline MoS$_2$ without relying on specific grain boundary topology. We further integrate these polycrystalline MoS$_2$ memristors with a top gate to achieve non-volatile memory cells with switching ratios as high as ~10$^4$.

## II. DEVICE FABRICATION

In this study, two device variants were fabricated, namely back-gated and top-gated MoS$_2$ memristors. In both cases, the first step is to grow a continuous polycrystalline film of monolayer MoS$_2$ by CVD (grain size of 3-5 μm) on 300 nm SiO$_2$-capped Si substrates [17] (Fig. 1). The resulting MoS$_2$ film is etched into rectangular channels by Ar-based reactive ion etching. Finally, source and drain electrodes (3 nm Ti/70 nm Au) are defined by photolithography, metal evaporation, and liftoff processes to obtain back-gated MoS$_2$ memristors with typical channel length (*L*) of 5-15 μm and width (*W*) of 50-100 μm (Fig. 2). Additionally, for top-gated memristors, a 30 nm thick Al$_2$O$_3$ gate dielectric layer is grown by atomic layer deposition (ALD) at 100 °C through photoresist patterns prior to the final step of top gate metal deposition. All charge transport measurements were performed in a vacuum probe station (pressure < 5 × 10$^{-5}$ Torr) connected to Keithley source measure units that are controlled by LabVIEW programs.

## III. RESULTS AND DISCUSSION

Current-voltage ($I_D$–$V_D$) characteristics of polycrystalline MoS$_2$ memristors show bipolar resistive switching, where the device starts in a high resistance state (HRS) for positive $V_D$ and gradually switches to a low resistance state (LRS) at high biases (sweeps 1 and 2 in Fig. 3). The device continues to stay in LRS at negative $V_D$ and then gradually resets to HRS at higher biases (sweeps 3 and 4 in Fig. 3). $I_D$ values in each HRS and LRS can be controlled with the gate bias $V_G$ (Fig. 4) by a factor of 10$^3$. Positive $V_D$ characteristics show the low saturation current of a reverse-biased Schottky diode at the source contact, whereas negative $V_D$ characteristics show the large and exponentially increasing current of a forward-biased Schottky diode. It should be noted that bias sweeps of only positive or only negative $V_D$ values result in significant reduction in the loop size, and a full 80 V to -80 V sweep is necessary to achieve reversible HRS and LRS switching in subsequent cycles. Moreover, resistive switching is gradual, and the on/off ratio between sweeps 1 and 2 increases with sweep bias range. Subsequent to high $V_D$ sweeps, $I_D$–$V_G$ transfer characteristics show significant shifts in threshold voltage $V_{th}$ (10 V to 20 V) between HRS and LRS with negligible hysteresis within one $V_G$ loop (Figs. 5 and 6). This resistive switching differs from the common $V_G$ hysteresis behavior of a conventional field-effect transistor (FET). This shift in the $I_D$–$V_G$ curves results in resistive switching, where the resistance in each state can be varied continuously by $V_G$ (Fig. 7). Consequently, the switching ratio increases in the sub-threshold regime due to larger band-bending near the contacts, at the expense of lower current values.

MoS$_2$ memristors were tested for endurance, reversibility, and stability of resistive switching. The devices show evolving changes in HRS and LRS resistance for the first few cycles (e.g., 5 cycles for the device in Fig. 8), and then resistance states stabilize to nearly constant values with a gradual increase in switching ratio with further cycling. This behavior is consistent with interface-based soft switching at the contacts instead of filament-based hard switching in metal oxide memristors [4,5]. Fig. 9a shows a schematic of resistive switching during high biases ($V_D \gg 0$ V) in the sub-threshold regime ($V_G - V_{th} < 0$ V) [18]. Large band-bending at the source contact acts as a bottleneck for the carriers, resulting in resistive switching from dynamic evolution of the Schottky barrier due to migration of sulfur vacancies in MoS$_2$ and/or charge trapping in the oxide dielectric. Therefore, the smaller ratio during LRS to HRS switching for negative $V_D$ can be explained by a smaller band-bending at the drain contact because the Fermi level is closer to the conduction band edge (i.e., $V_G - V_D = -|V_G| + |V_D| \sim V_{th}$) (Fig. 9b). Further evidence for the role of defect migration in these devices is provided by the fact that single-crystal flakes without grain boundaries do not show memristive switching, which is consistent with recent scanning transmission electron microscopy and *ab-initio* calculations that have shown that MoS$_2$ grain boundaries provide efficient pathways for defect motion [19-21]. The present device characteristics are also similar to previously reported gate-tunable MoS$_2$ memristors that possessed individual grain boundaries in the channel [16]. However, since the devices here consist of tens of such grain boundaries, the need for specific topology is removed, enabling improved device scalability and reproducibility.

Top-grated MoS$_2$ memristors achieve qualitatively different resistive switching (Figs. 10 and 11). In this case, the transfer characteristics in HRS and LRS show significantly larger switching ratios in excess of $10^4$ (Fig. 12). This switching behavior is similar to floating gate or metal-nitride-oxide-silicon (MNOS) non-volatile memory devices [18]. $I_D$–$V_G$ characteristics flatten at positive $V_G$ values due to significantly larger contact resistance in the top-gated geometry, and the devices also do not turn off completely in LRS due to large electron doping. These effects could originate from the additional thermal stress encountered during the 8 hours of ALD growth of Al$_2$O$_3$ at 100 °C. Since large Schottky barriers are desirable for resistive switching, we decreased the overall electron doping by modifying the liftoff duration in n-methyl-2-pyrrolidone at 80 °C from overnight to just 30 min. As desired, the turn-off voltage in the resulting devices increased from –8 V to 4 V (Figs. 12, 13). These devices then underwent a set of five $V_D$ bias sweeps with the transfer characteristics being recorded between each sweep (Figs. 13 to 17). Reversible HRS-LRS switching in $I_D$–$V_G$ (ratio ~50-500) occurs within the first sweep with minor resistance change in the subsequent sweeps. Compared to back-gated memristors, top-gated devices show smaller memristive loops for $V_D > 0$ V (Fig. 17). These differences can be explained by the increased role of charge trapping in the Al$_2$O$_3$ top gate dielectric that screens the top gate field, resulting in a shift in $V_{th}$. The memristive loop for $V_D < 0$ V remains appreciable due to the overall decrease in doping (larger $V_{th}$) in top-gated MoS$_2$ devices that results in large band-bending even at negative V$_D$ (i.e., $V_G - V_D < V_{th}$), thus making the drain contact susceptible to defect-induced band modulation.

Finally, we note that due to the low resistive switching ratio (~10-500) in these gate-tunable MoS$_2$ memristors, they are not ideal for high-performance ReRAM applications. Instead, the key innovation is in the continuous tunability of the resistive switching by the gate electrode that mimics realistic neural functions [8,22,23]. Moreover, interface-based switching is primarily focused on tunability of HRS and LRS states. This tunability of switching voltage can be achieved by integrating monolayer MoS$_2$ FETs with hard switching conventional memristors or by controlling grain boundary topologies in a monolayer MoS$_2$ memristor (Fig. 18) [4,17].

## IV. CONCLUSIONS

In summary, we have presented gate-tunable memristors based on CVD-grown polycrystalline MoS$_2$ films. The devices do not require specific topology of grain boundaries as shown in previous gate-tunable MoS$_2$ memristors [16]. Instead, the memristive characteristics are achieved by employing several grain boundaries in each device channel. Therefore, we anticipate that operating voltages can be decreased by proportional scaling of the channel dimensions, grain sizes, and gate oxide thickness. Resistive switching ratios of ~100 with gate tunability over ~1000x hold promise for emerging neuromorphic architectures [4,8,22,23]. The integration of memristor and transistor characteristics in a single device also presents opportunities for cross-point array architectures.

ACKNOWLEDGMENT

This research was supported by the National Science Foundation MRSEC (DMR-1121262) and 2-DARE program (EFRI-1433510). H.-S.L. also acknowledges the Basic Science Research Program of the National Research Foundation of Korea (NRF) that is funded by the Ministry of Education (2017R1A6A3A03008332).

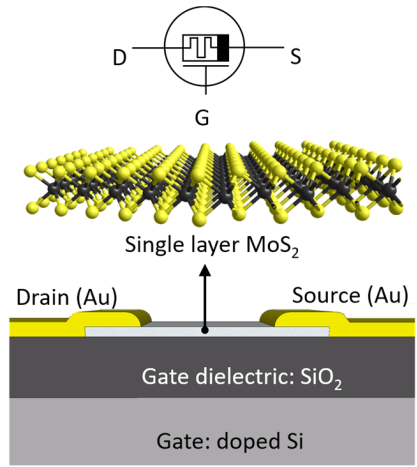

Fig. 1. Symbol and schematic of a gate-tunable memristor on CVD-grown polycrystalline monolayer $MoS_2$ on a thermal oxide-coated doped Si substrate that acts as a global bottom gate.

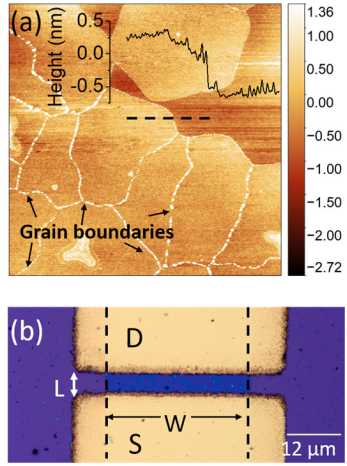

Fig. 2. (a) Atomic force microscopy image of polycrystalline $MoS_2$, revealing grain boundaries and monolayer thickness of 0.8 nm from the height profile along the dashed line. (b) Optical image of a $MoS_2$ memristor with channel length $L$ and width $W$.

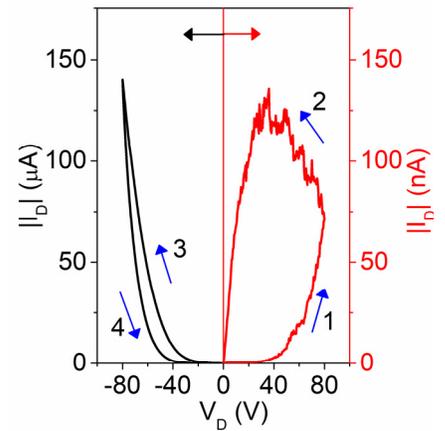

Fig. 3. Current-voltage characteristics ($I_D$-$V_D$) of a $MoS_2$ memristor where $V_D$ is swept in the order of 1-2-3-4. $I_D$ values for positive and negative $V_D$ are shown on the left (black) and right (red) vertical axes, respectively. Sweep rate = 5 V/sec. Gate voltage $V_G$ = 10 V.

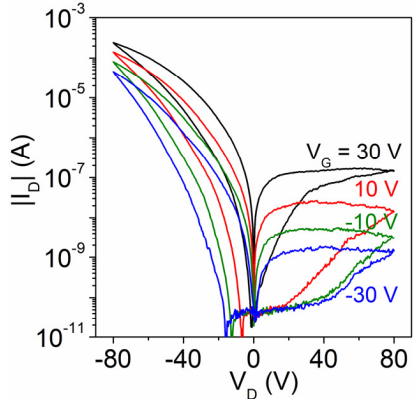

Fig. 4. $I_D$-$V_D$ characteristics of a $MoS_2$ memristor with $V_G$ varying from 30 V to -30 V. The on/off ratio is larger for $V_D > 0$ V and increases with decreasing $V_G$.

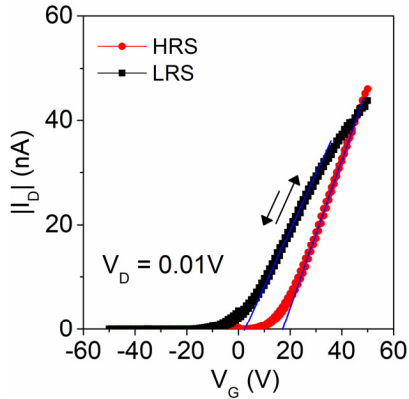

Fig. 5. Transfer characteristics ($I_D$-$V_G$) of a $MoS_2$ memristor in HRS and LRS states, showing a shift of 16 V in the threshold voltage ($V_{th}$) with minimal hysteresis.

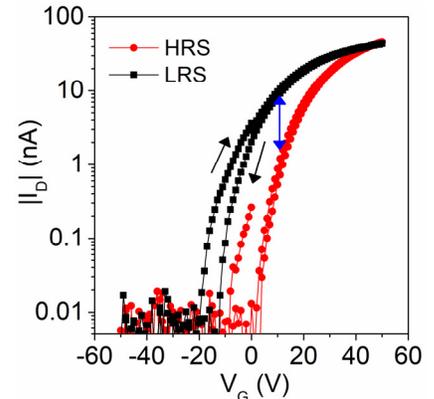

Fig. 6. Log-linear plot of the data from Fig. 5 with the switching ratio indicated by the blue arrow at $V_G$ = 10 V.

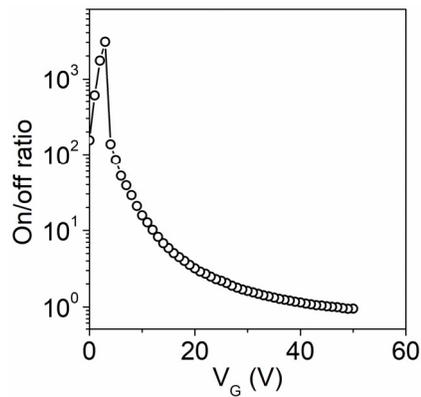

Fig. 7. Plot of the ratio of resistances in HRS and LRS at $V_D$ = 50 mV (on/off ratio) as a function of $V_G$.

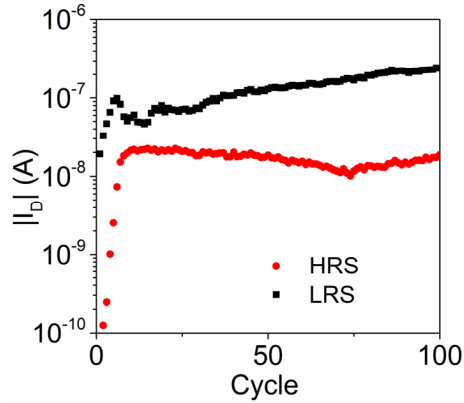

Fig. 8. Endurance characteristics of a gate-tunable memristor for 100 sweep cycles at $V_G$ = 30 V. The first five sweeps involve an extended electroforming process after which the resistance settles to a relatively constant value.

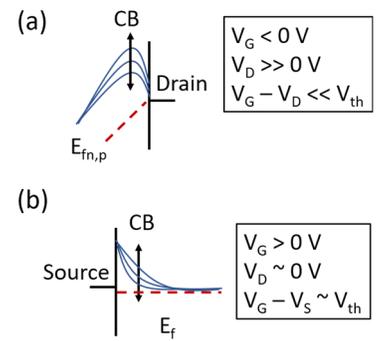

Fig. 9. Band diagram of a $MoS_2$ memristor near drain and source contacts (a) far from equilibrium, $V_D \gg 0$ V and (b) near equilibrium, $V_D \sim 0$ V. Memristive switching involves dynamic variation in the Schottky barrier due to defect migration near the contacts assisted by grain boundaries.

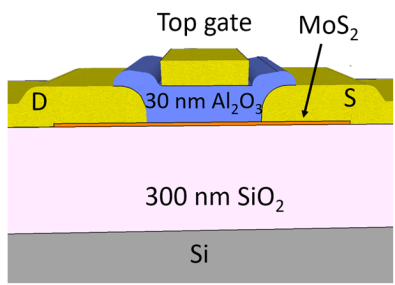

Fig. 10. Schematic of a top-gated MoS$_2$ memristor with 30 nm thick top gate Al$_2$O$_3$ dielectric grown by ALD.

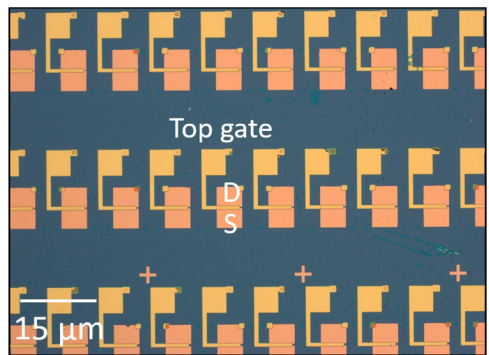

Fig. 11. Optical image of on an array of top-gated MoS$_2$ devices. The global back gate Si is grounded throughout this study.

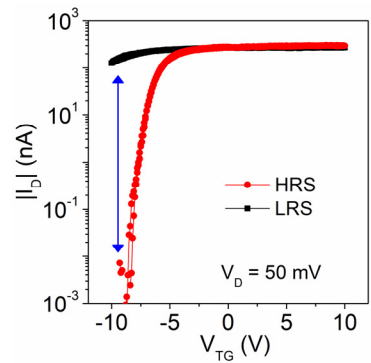

Fig. 12. $I_D$-$V_{TG}$ characteristics of a top-gated MoS$_2$ memristor that is switched between HRS and LRS by a 1 V $V_D$ pulse to produce an on/off ratio > $10^4$.

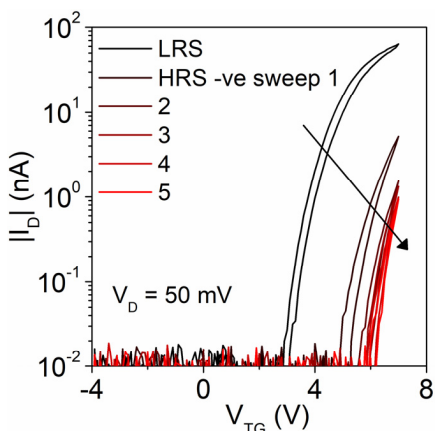

Fig. 13. $I_D$-$V_{TG}$ characteristics of a top-gated memristor where MoS$_2$ is doped to enhancement mode. Transition to HRS is shown over five successive negative (-ve) $V_D$ sweeps, as shown in Fig. 16. The resistance increases by >50x in this case.

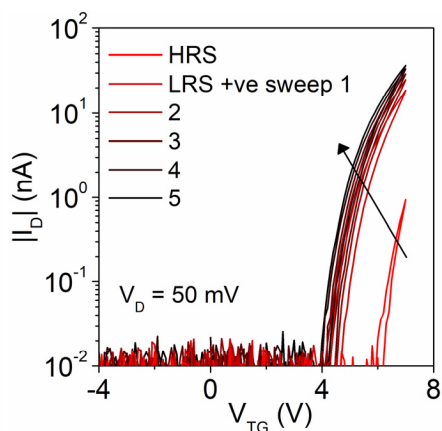

Fig. 14. $I_D$-$V_{TG}$ characteristics of the device from Fig. 13 after positive (+ve) $V_D$ sweeps, as shown in Fig. 17. The original LRS of the device is recovered.

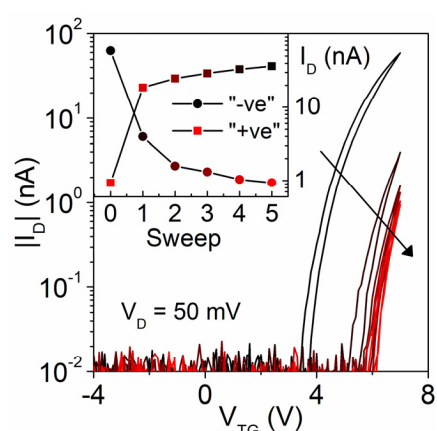

Fig. 15. $I_D$-$V_{TG}$ characteristics of the same device after repeated negative sweeps (Fig. 16), showing reversibility of HRS-LRS switching. Inset shows $I_D$ values after each sweep at $V_G$ = 7 V, resulting in neuromorphic potentiation and depression for positive and negative sweeps, respectively.

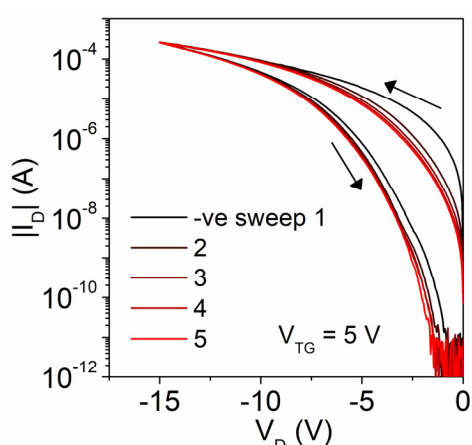

Fig. 16. $V_D$ sweeps of -15 V are used to switch the top-gated MoS$_2$ device from LRS to HRS (Figs. 13 and 15). Sweep characteristics show memristive switching and exponential dependence resulting from the forward-biased Schottky diode at the source contact (Fig. 9).

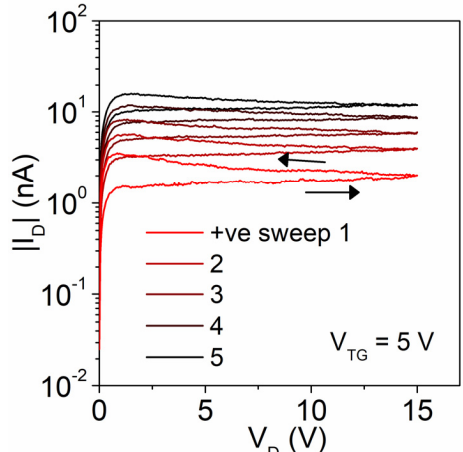

Fig. 17. $V_D$ sweeps of 15 V are used to switch the device from HRS to LRS (Fig. 14). Positive sweep characteristics resemble current saturation of a reverse-biased Schottky diode that increases with each sweep without a large memristive loop.

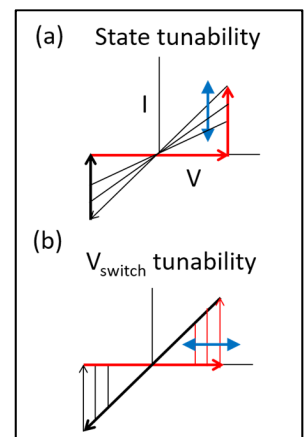

Fig. 18. (a) Conceptual schematic of gate tunability of resistance states in interface-based soft switching memristors. (b) Schematic of tunability of switching voltage in filament-based hard switching memristors.